# Visualizing the Zhang-Rice singlet, molecular orbitals and pair formation in cuprate

Shusen Ye[1], Jianfa Zhao[2], Zhiheng Yao[1], Sixuan Chen[1], Zehao Dong[1], Xintong Li[2], Luchuan Shi[2], Qingqing Liu[2], Changqing Jin[2], and Yayu Wang[1,3,4*]

[1]*State Key Laboratory of Low Dimensional Quantum Physics, Department of Physics, Tsinghua University, Beijing 100084, P. R. China*

[2]*Beijing National Laboratory for Condensed Matter Physics, Institute of Physics, Chinese Academy of Sciences, Beijing 100190, P. R. China*

[3]*New Cornerstone Science Laboratory, Frontier Science Center for Quantum Information, Beijing 100084, P. R. China*

[4]*Hefei National Laboratory, Hefei, 230088, China*

*Corresponding author. Email: yayuwang@tsinghua.edu.cn

**The parent compound of cuprates is a charge-transfer-type Mott insulator with strong hybridization between the Cu $3d_{x^2-y^2}$ and O $2p$ orbitals[1]. A key question concerning the pairing mechanism is the behavior of doped holes in the antiferromagnetic (AF) Mott insulator background, which is a prototypical quantum many-body problem[2]. It was proposed that doped hole on the O site tends to form a singlet, known as Zhang-Rice singlet (ZRS), with the unpaired Cu spin[3]. But experimentally little is known about the properties of a single hole and the interplay between them that leads to superconductivity. Here we use scanning tunneling microscopy to visualize the electronic states in hole-doped $Ca_2CuO_2Cl_2$, aiming to establish the atomic-scale local basis for pair formation. A single doped hole is shown to have an in-gap state and a clover-shaped spatial distribution that can be attributed to a localized ZRS. When the dopants are close enough, they develop delocalized molecular orbitals with characteristic stripe- and ladder-shaped patterns, accompanied by the opening of a small gap around the Fermi level ($E_F$). With increasing doping, the molecular orbitals proliferate in space and gradually form densely packed plaquettes, but the stripe and ladder patterns remain nearly the same. The low-energy electronic states of the molecular orbitals are intimately related to the local pairing properties, thus play a vitally important role in the emergence of superconductivity. We propose that the Cooper pair is formed by two holes occupying the stripe-like molecular orbital, while the attractive interaction is mediated by the AF spin background.**

In the resonating valence bond (RVB) theory for high $T_c$ superconductivity, Anderson pointed out that the cuprates are doped Mott insulators, and the pairing originates from the AF superexchange between Cu spins[1]. The essential physics of RVB can be captured by the Hubbard model[4], which involves the hopping of charges on a periodic lattice with strong onsite Coulomb repulsion. In the past few decades, the Hubbard model has become one of the most thoroughly investigated theoretical models in physics, but there is still controversy whether the plain Hubbard model or its variants can generate Cooper pairs[5–8]. This situation is in stark contrast to the fact that high $T_c$ superconductivity in cuprate is a remarkably robust state of matter insensitive to material details.

A factor that is certainly missing in the Hubbard model is the randomly distributed local potential induced by charge dopants. It is either considered an unimportant detail, or is omitted to avoid further complication. However, in a "doped" Mott insulator the dopants are the source of novel physics, thus must be treated properly in a realistic model. This point is eminently exemplified in doped semiconductors, in which a charge dopant is modeled as a hydrogen-like (H-like) atom[9]. The doped charge forms a bound state with the dopant ion by screened Coulomb potential, thus its energy level and spatial wavefunction can be calculated quantitatively. Instead of being a nuisance, the dopant plays an indispensable role in fundamental physics such as metal-insulator transition[10], Anderson localization[11] and quantum Hall effect[12].

Similarly, a monovalent hole donor in cuprate can be viewed as a H-like atom embedded in the AF Mott insulator. When two dopants are close to each other, they may develop extended molecular orbitals due to the overlap of wavefunctions. With sufficiently dense dopants, the doped holes will pair up and condense into the superconducting (SC) state. Mapping out the electronic structure of each stage is of paramount importance for elucidating the pairing mechanism, but experimental effort along this line of attack has been inadequate. There was a theoretical proposal of molecular-orbital basis for superconductivity[13], but it was before the high $T_c$ era and focused on metallic systems with dynamic Jahn-Teller coupling. The goal of this work is to construct the atomic and molecular orbital basis for pair formation in cuprate from the doped Mott insulator perspective.

Scanning tunneling microscopy (STM) is a powerful technique to study doped Mott insulator because the physics is intrinsically local. An ideal material system for our purpose is alkali metal doped $Ca_2CuO_2Cl_2$ (CCOC). As shown by the schematic structure in Fig. 1a inset, the crystal cleaves easily between two adjacent Cl layers with weak van der Waals bonding, making it suitable for surface probes such as angle-resolved photoemission spectroscopy[14–17] and STM[18–22]. Each alkali element substitution of Ca donates a single hole into the $CuO_2$ plane, thus is a perfect realization of H-like atom in an AF Mott insulator. Figure 1a shows a large topographic image of CCOC with dilute hole donor density $p = 0.03$ taken at $T = 77$ K, which presents an overview of the randomly distributed dopants. The surface Cl atoms constitute a square lattice with lattice constant $a_0 = 3.9$ Å, and the dark spots are sporadic Cl vacancies. A dopant can be identified by the bright Cl tetramer surrounding the Ca site, and the Cu sites in the $CuO_2$ plane lie directly underneath the surface Cl atoms.

To study the electronic state of a single doped hole, in Fig. 1b we zoom into a relatively isolated dopant (marked by the magenta dot) at $T = 23$ K. Figure 1c displays three $dI/dV$ curves at representative locations designated in Fig. 1d, the density of state (DOS) map taken at bias voltage $V_b = 400$ mV (see Supplementary Material SI I for details), which exhibits a unique 'four-lobe' clover pattern surrounding the central Cu site. The blue curve in Fig. 1c at a defect-free area reveals the charge transfer gap with size around 2.0 eV between the charge transfer band (CTB) and upper Hubbard band (UHB) of a pristine Mott insulator[21]. The black and red curves taken at the clover center and lobes both have a broad in-gap state around $V_b = 0.6$-$0.7$ V, and the CTB at negative bias shifts towards $E_F$ due to local hole doping. The spectral weight of the UHB above $V_b = 1.8$ V is suppressed and transferred to the in-gap state, which is a characteristic behavior of doped Mott insulator[21,23]. The series of $dI/dV$ spectra in Fig. 1f along the line in Fig. 1d clearly shows the shift of in-gap state peak and weight to lower energy on the clover lobes. These features are consistent with the ZRS model schematically depicted in Fig. 1e. To optimize the interaction between the unpaired Cu spin and doped hole spin, the four O hole wavefunctions in a $CuO_4$ ligand combine into the $d_{x2-y2}$ symmetry to maximize the overlap with central Cu $3d_{x2-y2}$ orbital. But they have destructive interference upon reaching an

*s*-wave tip at the clover center, resulting in a suppression of DOS. In Supplementary Material SI II we give a detailed analysis of the internal phase relation and tunneling path of a ZRS state, and generate a simulation that agrees well with the observed clover pattern. The central Cu site is selected by the potential landscape of the environment, which breaks the degeneracy of four otherwise equivalent Cu sites neighboring the dopant. The doped hole state is Cu-centered rather than dopant-centered, which is in sharp contrast to other impurity-induced electronic states in cuprates[24–29]. The dopant here only provides a weak Coulomb attraction that loosely binds the doped hole, but does not participate in the creation of ZRS state directly.

Figure 2a displays the topography of a larger field of view (FOV) on the same sample at $T = 23$ K, and the DOS maps (Fig. 2b-c) reveal several highly intriguing features. The dark areas have no DOS at all in the energy range of interest, representing the Mott insulator background without hole doping. The DOS map at $V_b = 400$ mV (Fig. 2c) reveals several clover-shaped ZRSs bound by the respective hole dopants marked by magenta dots. For hole dopants lying close to each other, clusters or chains of electronic states with varied shapes start to emerge. The low energy ($V_b = 50$ mV) DOS map in Fig. 2b reveals arrays of bright stripes running parallel to the Cu-O bond direction between neighboring hole dopants. At high energy ($V_b = 400$ mV), the dominant features are dense ladder rungs aligned along the Cu-O bond direction perpendicular to the stripes in Fig. 2b. Both the shape and orthogonality of the stripe and ladder patterns are evident manifestations of two types of molecular orbital[30] formed by the combination of atomic wavefunctions between neighboring hole dopants. The main features are highly reminiscent of the drastically different bonding and antibonding molecular orbitals observed in real molecules such as pentacene[31], phenazine[32] and graphene nanoribbons[33]. To extract the periodic features, in Fig. 2h-i we perform Fourier transform (FT) of the DOS maps. The linecut along the lattice direction in Fig. 2h reveals a broad range of FT wavevectors between $0.70 \times (2\pi/a_0)$ and $0.90 \times (2\pi/a_0)$ (Fig. 2j), indicating the existence of real-space pattern with short-range periodicity from $1.1a_0$ to $1.4a_0$. There also exists a shallow peak around $0.25\times(2\pi/a_0)$, corresponding to a spatial pattern with periodicity $\sim 4a_0$.

We then zoom into a small area enclosed by the dashed square in Fig. 2b with a quasi-one-dimensional (1D) array of stripe-like molecular orbitals. The DOS maps at $\pm 50$ mV (Fig. 2d-e) are quite similar to each other, demonstrating approximate particle-hole symmetry for the electronic states closest to $E_F$. At intermediate energy scale (Fig. 2f), the main features have squarish shape with short stripes straddling both Cu-O bond directions, which then crossover to the ladder rungs at higher energy (Fig. 2g). The line profile in Fig. 2k reveals that although the central stripe lies on top of the Cu-O bond, the two side stripes are located at $\sim 1.2a_0$ away, thus are incommensurate with the lattice marked by yellow dots. Such distances between neighboring stripes, which also exist in other stripe-like molecular orbitals in the FOV, are responsible for the broad peak around $0.80\times(2\pi/a_0)$ in Fig. 2j. The lattice wavevector in the FT map at 50 mV (Fig. 2h and 2j) is nearly invisible, reinforcing that the stripes are essentially incommensurate. The line profile in Fig. 2l along the stripe array reveals nodes between adjacent stripe segments with distance $\sim 4a_0$. The approximately $4a_0$ periodicity also exists in other stripe arrays with different dopant configurations in the FOV in Fig. 2b, and together they contribute to the FT peak around $0.25\times(2\pi/a_0)$ in Fig. 2j. The high-energy ladder pattern in Fig. 2g is largely commensurate with the lattice, which is also confirmed by the obvious lattice wavevector in the FT intensity in Fig. 2j.

The molecular orbitals not only have characteristic spatial features, but also create novel electronic structures. In Fig. 3 we present the topography, DOS maps, and d$I$/d$V$ spectra on a cluster of dopants in the $p = 0.03$ sample at $T = 23$ K. The dashed red rectangle in Fig. 3a contains several dopants on a row, and the DOS maps at $V_b = 50$ mV (Fig. 3b) and 400 mV (Fig. 3c) exhibit obvious stripe- and ladder-shaped molecular orbitals. Figure 3e displays the d$I$/d$V$ spectra taken on several representative locations marked in the inset with corresponding colors. Compared to that on the ZRS, the in-gap state peak moves to lower energy towards $E_F$, and are typically around $V_b$ = 350 mV. More interestingly, a gap-like feature around $E_F$ with a shoulder ~ 20 mV (indicated by the black arrow) starts to emerge on bright stripes. Figures 3e and 3f are a series of d$I$/d$V$ spectra taken along two perpendicular lines, which demonstrate that the gap-like features are closely associated with stripe-like molecular orbital. The small gap is still in the preliminary stage thus is not full-fledged, but it represents the first spectroscopic feature near $E_F$ with an energy scale comparable to the SC gap size. It signifies the essential role of the stripe-like molecular orbital on the genesis of low-energy electronic states, which are essential prerequisites for the emergence of superconductivity.

Next, we move to a CCOC with higher hole density $p$ = 0.07, which is still in the insulating regime but close to the insulator-superconductor phase boundary[34]. The sample becomes less insulating, which allows us to perform STM measurements at $T$ = 5 K. The topography in Fig. 4a now has apparent stripy features along both Cu-O bond directions, leading to a crisscross network occupying the majority of the FOV. Figure 4b displays the DOS map at 30 mV taken in the FOV enclosed by the red square in Fig. 4a, which reveals three types of regions: dark pits without hole doping, small patch of squarish plaquettes, and quasi-1D stripe arrays. The DOS map at 300 mV (Fig. 4c) is covered with rather dense ladders that are orthogonal to the stripes. Compared to that in the $p$ = 0.03 sample, the spatial occupancy of molecular orbital grows significantly, but the characteristic local patterns are nearly the same. Figure 4e displays the d$I$/d$V$ curves obtained at several locations marked in Fig. 4b. A common feature is the existence of a small gap around $E_F$ in all curves outside the dark pits, but there are also important differences in terms of gap size and lineshape. The golden and magenta curves on the bright long stripe have a V-shaped gap with size ~ 10 meV, accompanied by a pseudogap with size ~ 150 meV (indicated by the black arrow). The overall features are highly analogous to that in previous report on SC CCOC[19]. This is not unexpected because as discovered recently[35], insulating cuprates near the SC phase boundary host abundant local Cooper pairs with short-range phase correlation. A more surprising finding is the black curve taken on a relatively isolated short stripe, which exhibits a U-shaped gap with edges at ±15 meV. The low-energy spectra in Fig. 4f shows that the gap is flat at the bottom with two well-defined peaks at the edges. The brown and red spectra have intermediate lineshapes between typical U- and V-shaped gaps. To illustrate the emergence of low-energy gap, we apply clustering analysis based on the tunneling asymmetry defined as $Z = \int_{200\text{ mV}}^{300\text{ mV}} \text{DOS}(E)\text{d}E \,/\, \int_{-300\text{ mV}}^{-200\text{ mV}} \text{DOS}(E)\text{d}E$, which reflects how far the system is doped away from the Mott insulator[36]. As shown in Fig. 4d, with decreasing tunneling asymmetry, the low-energy gap comes into form and evolves into well-defined V-shaped gap, accompanied by the enhancement of pseudogap features[19]. Because the U-shaped gap only exists on a small fraction of the FOV, its feature is smeared out.

Figure 5 displays the zoomed-in topography, DOS maps and d$I$/d$V$ spectra on the two areas enclosed by the dashed boxes in Fig. 4b. The stripe and ladder patterns in Fig. 5b-c and Fig. 5f-g are strikingly similar to that shown in Fig. 3b-c for the $p$ = 0.03 sample, indicating the common origin of molecular orbitals. Figures 5d and 5h display the spectral linecuts in Fig. 5b and 5f, respectively, demonstrating that in both places the small gap feature is associated with the molecular orbital and mainly exist in places with evident low-energy stripe pattern. Moreover, at locations without well-established pseudogap feature, the small gap is also severely under-developed. Despite these common trends, there are also important differences between these two sets of spectra. Especially, the spectroscopic gap on the left stripe typically has a V-shape, whereas on the right stripe the gap always has a U-shape. It is puzzling why a nodeless gap exists in this regime of cuprate, while just a few nanometers away the gap has the expected V-shape for $d$-wave pairing symmetry. We can gain some insights from theoretical proposals in stripe-based mechanism, which envisages that the pairing symmetry is sensitive to the coupling between neighboring stripes or ladders[37]. The stripe with V-shaped gap is more extended in both the longitudinal and perpendicular directions with more regular $4a_0$ periodicity, thus bears stronger resemblance to the checkerboard order with *d*-wave pairing symmetry and V-shaped SC gap at higher dopings[18,19]. In contrast, the region with U-shaped gap has rather irregular stripe structure and weaker coupling between neighboring stripes within the array. Because it is quite far from the two-dimensional (2D) limit, the *d*-wave symmetry is ill-defined and hence the gap has a nodeless U-shape, which may be related to proposals based on strongly correlated stripe or lattice models[37–40].

The pronounced stripy features in the topography of the $p = 0.07$ sample is surprising at the first glance because the distribution of dopants is random, but it can be explained readily from the molecular orbital viewpoint. As well-established in quantum chemistry, the three key requirements for molecular orbital formation are symmetry, equal energy, and overlap of wavefunctions[30]. These conditions are best satisfied when neighboring dopants are situated along the same row of Cu-O bonds. The formation of stripe-like molecular orbitals is thus a self-selective process because the doped hole preferably finds a nearby partner with relative orientation closest to the Cu-O bond direction. As a consequence, the stripe-like molecular orbital becomes the dominant conduction channel for doped holes at energies around $E_F$, which in turn gives rise to the stripy topography. We note that the origin and properties of the stripe-like molecular orbital are fundamentally different from the stripe order derived from the Hubbard and *t-J* models, in which doped holes self-assemble into 1D chains separating the AF spin domains[41–44]. The stripe-like molecular orbital observed here is not a spontaneous $C_4$-to-$C_2$ symmetry breaking electronic state because the dopants already set a preferred local orientation. Moreover, the quasi-1D stripe wavefunction here is not exactly located on the atomic or bond site, as assumed in most lattice models, which may introduce new type of interaction between doped holes and local spins[41,42,45–47]. Despite these differences, the extensive theoretical studies on the stripe model may be exploited to address certain aspects of the physics here, such as the pairing symmetry discussed above.

Lastly, we investigate a SC sample with $T_c$ = 14 K and $p$ = 0.10 (ref. 34,48). The topography in Fig. 6a and the DOS maps in Fig. 6b-c are generally consistent with previous results on similar CCOC[18,19]. The DOS map at 30 mV shows well-formed plaquettes with internal stripe structure, whereas the DOS map at 200 mV consists of complicated maze-like patterns. Locally the spatial features are nearly identical to the stripe and ladder molecular orbitals in the dilute doping regime, but now the plaquettes are densely packed and give rise to the renowned checkerboard order. The FT of the DOS map at 30 mV (Fig. 6e) reveals two prominent features, previously assigned as $(1/4)\times(2\pi/a_0)$ and $(3/4)\times(2\pi/a_0)$ wavevectors, corresponding to the checkerboard order with periodicity $4a_0$ and internal stripe structure[18,35]. However, the linecuts in Fig. 6d indicate that instead of two sharp FT peaks, there is a broad range of wavevectors around $0.28\times(2\pi/a_0)$ and $0.72\times(2\pi/a_0)$. A closer examination of previous results on Na doped CCOC (Fig. 2g-i in ref. 18) leads to a similar conclusion. The departure from commensurate $(1/4)\times(2\pi/a_0)$ wavevector is corroborated by bulk-sensitive X-ray scattering result on the same $T_c$ samples[49], which ensures that the surface charge order reflects the bulk property. The incommensurate checkerboard could be attributed to the random distribution of plaquettes with slightly varied shapes and sizes, thus are not long-range order with a well-defined periodicity. Given the new insights from molecular orbitals, now we can give a natural explanation for these perplexing textures. Figure 6g summarizes the statistical histogram of inter-stripe distance within a plaquette in Fig. 6b (see Supplementary Material SI III for details), which has a broad distribution around $\sim 1.3a_0$ hence corresponds to the series of wavevectors around $0.72\times(2\pi/a_0)$. The FT in Fig. 6d reveals that at higher energy, the spatial pattern becomes more commensurate with lattice, which is also consistent with the ladder-type molecular orbital in Fig. 2c. The new feature here is an arc of wavevectors as marked by the magenta arc in Fig. 6e, which can be attributed to the slight wiggling of the ladders. Therefore, the peculiar spatial features can be well described as a mosaic of plaquettes with slightly varied sizes and incommensurate molecular orbitals on the CuO$_2$ plane.

To further confirm the molecular orbital origin of the charge order in the SC phase[18,20,46,50], we provide two quantitative analysis. The first one is the correlation between the checkerboard order at low energy and ladder pattern at higher energy. We define a local orientational order parameter by the difference of modulation amplitude along two Cu-O bond directions, $\beta(\mathbf{r}, E) = A_{Q_x}(\mathbf{r}, E) - A_{Q_y}(\mathbf{r}, E)$, where $A_{Q_{x,y}}(\mathbf{r}, E)$ is the local amplitude of lattice wavevector $Q_x$ or $Q_y$ calculated by spatial lock-in technique[51,52] on the corresponding DOS map. As shown in Fig. 6h, $\beta(\mathbf{r}, 30\,\mathrm{mV})$ is positively correlated with $\beta(\mathbf{r}, -30\,\mathrm{mV})$ and negatively correlated with $\beta(\mathbf{r}, 300\,\mathrm{mV})$. These trends perfectly match that of isolated molecular orbitals in the $p$ = 0.03 sample (Fig. 2d-g), where the stripe-like orbitals at ±50 mV are particle-hole-symmetric but are perpendicular to the ladder-like orbitals at +300 mV. The second one is the correlation between the internal stripe orientation within each plaquette and the inter-plaquette spatial distribution. As described in Supplementary Material SI IV, we locate every plaquette and the internal stripe orientation for the FOV in Fig. 6b. Starting from each plaquette center, we count the relative orientation number defined as $n_{\mathrm{para}} - n_{\mathrm{perp}}$, where $n_{\mathrm{para(perp)}}$ is the number of plaquette parallel (perpendicular) to the stripe in the original plaquette. As shown by the statistics in Fig. 6i, there are pronounced peaks slightly below 4$a_0$ along both Cu-O directions, while the 4$a_0$ correlation is much stronger along the stripe direction than the perpendicular direction. This feature is also closely connected to the isolated stripe-shaped molecular orbital in the $p$ = 0.03 (Fig. 2g), which display arrays of 4$a_0$ segments aligned along the internal stripes. These strong similarities provide unambiguous evidence that the spatial characteristics of the SC phase can be described by the dense packing of dopant-induced molecular orbitals, or electronic liquid crystal phase of shaped molecule with intrinsically 1D nature[46,53].

The molecular orbitals not only account for the spatial patterns, but also have profound implications on the pair formation process. Firstly, there exists a close correlation between the spectroscopic features of superconductivity with the local stripy patterns. The series of d$I$/d$V$ curves in Fig. 7c-d and g-h reveal that the SC coherence peaks are more pronounced and the SC gaps are deeper on top of the bright stripes, indicating the enhanced local pair density on the stripe-like molecular orbitals[54,55]. These features are qualitatively similar to that in the $p$ = 0.03 and 0.07 samples, which indicate that the SC gap derives from the stripe-like molecular orbitals of doped holes. The positive correlation between the SC properties and low-energy stripe patterns have been observed before in $Bi_2Sr_2CuO_{6+\delta}$ (Bi-2201) with various doping levels[35,55]. Secondly, by counting the number of plaquette in Fig. 6b, we estimate that on average each plaquette contains approximately two holes given the hole density $p$ = 0.10 (see Supplementary Material SI V for details). Each two-hole plaquette with a size around 4$a_0$ is thus the fundamental unit for Cooper pairing. This conclusion has also been drawn before on severely underdoped Bi-2201 (ref. 35) before the notion of molecular orbital. To further strengthen this statement, we show in Fig. 7a-b a region (the magenta dashed box in Fig. 6b) with a dark pit with regular rectangle shape and size ~4$a_0$×8$a_0$, corresponding to two connected plaquettes. The d$I$/d$V$ curves taken along the line indicates that at the center of the dark pit the SC gap is absent, and the topography in Fig. 6a reveals the absence of dopant in this small area. Another less obvious evidence is the existence of wavevector near 0.14×(2π/$a_0$) in the FT linecut in Fig. 6d, which corresponds to a real-space length scale ~7$a_0$. In Supplementary Material SI V we prove that this is due to the existence of numerous dark pits with typical size around 3-4$a_0$, so that the plaquettes surrounding the pits have an average distance around ~7$a_0$.

A likely reason why most plaquettes, or two-hole molecules, have a typical size around (or slightly smaller than) $4a_0$ is because the delocalization of holes can save kinetic energy, but, as a consequence, the suppression of AF order costs superexchange energy. The competition between these two factors leads to a Lennard-Jones-type energy landscape with a characteristic length scale around $4a_0$ corresponding to the potential minimum. The plaquettes tend to form a periodic pattern plausibly because the stripe-shaped molecular orbital acts as a quasi-1D conducting channel in the correlated background with divergent charge density wave susceptibility[45,56,57]. The internal structure of each plaquette has various shapes such as double-stripe, triple-stripe and small squares, because the molecular orbitals are affected by the local configuration of hole dopants. The polyatomic configuration along the Cu-O bond direction favors stripe-like patterns, whereas that along the diagonal direction may generate squarish patterns. Although locally the $C_4$ symmetry is reduced to $C_2$ by stripe-like molecular orbitals in many places, globally the $C_4$ symmetry is recovered because statistically there is no preferred stripe orientation due to the random distribution of dopants. This is consistent with the $d$-wave SC pairing symmetry and the momentum-space electronic structure in the normal state[58,59].

The molecular orbitals observed here provide a vivid revelation of the true meaning of "doping" in an AF Mott insulator. The doped holes in cuprate do not hop on a periodic lattice with AF order and onsite Coulomb repulsion, as conceived in the many-body lattice models. At the dilute limit, a single doped hole is weakly bound to the dopant, forming a H-like atom with ZRS state. For short enough inter-dopant distance, the combination of single hole state leads to delocalized molecular orbitals, which proliferate with progressive doping and gradually form densely packed plaquettes. The molecular orbital picture gives a natural and unified explanation for the complex real-space electronic structures in cuprates, including the microscopic inhomogeneity[60,61], mesoscale phase separation[20,23,35], and macroscopic charge order[18,49]. In the lightly doped regime, clusters of molecular orbitals immersed in the Mott insulator background generate strong spatial inhomogeneity and phase separation. With increasing doping, the molecular orbitals congregate into an interconnected framework that appears like a checkerboard order at low-energy and ladder pattern at higher energy. But as discussed above, these features can be most accurately described as a mosaic of plaquettes with internal stripe-like molecular orbital. These patterns do not compete with superconductivity as suspected previously, but are instead the prerequisite backbone for pair formation because the low-energy electronic states first appear on the stripes.

The stripe-like molecular orbital plays a pivotal role on superconductivity, which prompts a nearly real-space pairing picture in the BCS to BEC crossover regime[62]. In the SC state, a two-hole molecular plaquette has a typical size around $4a_0$, hence local hole density $p \sim 0.125$. At such hole density, the AF order of the underlying $CuO_2$ plane is destroyed[63], and the local Cu spins may behave like a genuine RVB quantum spin liquid. This two-component picture resolves the conundrum that both the itinerant and localized states in cuprates originate from the same Cu $3d_{x2-y2}$ orbital. Here we show that the itinerancy comes from the molecular orbitals of doped holes, whereas the local Cu spins are inherited from the parent Mott insulator. The doped holes are the component that eventually form Cooper pairs, which explains why the superfluid density is proportional to the doped hole density $p$ rather than the total hole density $1+p$ (ref. [64]). The hole conduction via stripe-like molecular orbitals and the scattering with local spin background may also help understand the anomalous "normal" state of cuprate, known as the strange metal phase[47,65,66].

The small energy gap around $E_F$ with size comparable to the SC gap first emerges and gradually develops predominantly on the stripe-like molecular orbitals, even in the lightly doped insulating regime. For a relatively isolated stripe, the small gap has a nodeless U-shape, which is consistent with the expectation for quasi-1D pairing. For coupled stripe arrays approaching the quasi-2D regime, the small gap evolves to the expected V-shape for *d*-wave symmetry. With further increase of hole density hence molecular orbital occupancy, bulk superconductivity is induced by the establishment of global phase coherence between localized Cooper pairs[35]. Below we propose two possible pairing mechanisms originated from the interplay between stripe-like molecular orbital and local AF spin background. The first one envisages a sizable exchange coupling $J_{hd}$ between the itinerant hole spin and local Cu spin due to the significant wavefunction overlap between the molecular orbital and the Cu $d_{x^2-y^2}$ orbital. Two doped holes with opposite spins in the stripe-like molecular orbital can form a Cooper pair mediated by the RVB spin singlet, which may be called "super-superexchange" pairing. Another possibility is the hopping of a pair of holes occupying the stripe-like molecular orbital to the spin singlet background, or to neighboring stripes to reduce kinetic energy. This picture is analogous to the spin gap proximity effect and pair hopping process proposed by the striped superconductivity model[45]. In these pictures, the small SC gap size ~10 meV characterizes the pairing strength between two holes, while the pseudogap is likely related to the superexchange $J_{dd} \sim 130$ meV between the nearest neighbor Cu spins[2,67].

The molecular orbital basis for pair formation visualized here provides fundamentally new insights for unveiling the mystery of high $T_c$ superconductivity in cuprates. The randomly distributed dopants break the lattice translation symmetry, and appear to exacerbate the challenge of solving the Hubbard model. However, we show that quite oppositely, the weakly bound ZRS creates a well-defined local basis for constructing the electronic states of doped holes in Mott insulator. It effectively transforms a strongly correlated many-body problem to a local few-body problem, which is more tractable computationally and amenable experimentally. In this scenario, superconductivity is a truly emergent phenomenon growing in a bottom-up manner from the proliferation of molecular orbitals of doped holes in real space. By forming incommensurate molecular orbitals lying at a short distance very close to the Cu spins, the doped holes achieve an optimized situation that avoids strong onsite Coulomb repulsion for motion, but can still take full advantage of the AF superexchange for pairing. The incommensurability of molecular orbitals also eliminates the propensity towards charge or spin orders in the lattice models[5–8], which usually compete with superconductivity and make the theoretical conclusions vulnerable. Besides, the AF superexchange $J_{dd}$ is among the largest due to strong hybridization between the Cu $3d_{x2-y2}$ and O $2p$ orbitals. It is the combination of these unusual attributes in one material system that makes the cuprate the best ambient pressure superconductor so far.

**Acknowledgement:** We thank Dung-Hai Lee, Zheng-Yu Weng, Tao Xiang, Shuo Yang, Fuchun Zhang, Guangming Zhang, Xingjiang Zhou, Bangfen Zhu and Liujun Zou for helpful discussions. The work at Tsinghua was supported by the Basic Science Center Project of NSFC (No. 52388201), the Innovation Program for Quantum Science and Technology (No. 2021ZD0302502), the Beijing Innovation Center for Future Chips, and the New Cornerstone Science Foundation through the New Cornerstone Investigator Program and the XPLORER PRIZE. The work at IOP was supported by the Young Elite Scientists Sponsorship Program of CAST (No. 2022QNRC001), the NSFC (No. 12204515) and MOST of China (No. 2021YFA1401800, 2022YFA1403804).

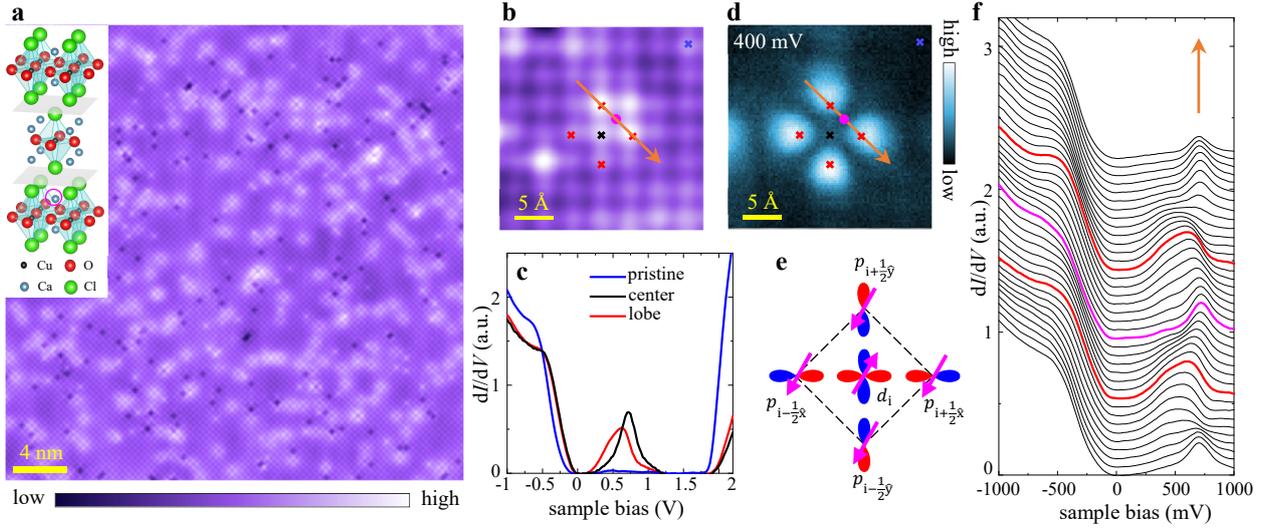

**Figure 1 | Topography of CCOC with $p$ = 0.03 and electronic structure of a Zhang-Rice singlet. a,** Large-scale topographic image of a $p$ = 0.03 CCOC taken at $T$ = 77 K. Inset: The schematic crystal structure of CCOC with the gray plane indicating the cleavage surface and magenta circle indicating the Ca site for alkali metal substitution. **b,** Zoomed-in topography of a relatively isolated hole donor taken at $T$ = 23 K with the dopant location marked by magenta dot. **c,** The d$I$/d$V$ curves taken on the pristine Mott insulator, center and lobe Cu sites of the clover-shaped electronic state, respectively. **d,** The DOS map of the in-gap state at 400 mV in the same FOV as **b**, displaying a four-lobe (red cross) clover pattern centered around a Cu site (black cross). **e,** The schematic diagram of a ZRS hosted by the CuO$_4$ ligand. The blue and red colors of the Cu 3$d$ and O 2$p$ orbitals indicate the relative phase of wavefunction. **f,** The d$I$/d$V$ spectra taken along the arrow in **b** and **d**. The spectral weight of in-gap state moves to lower energy on the clover lobes (red curves).

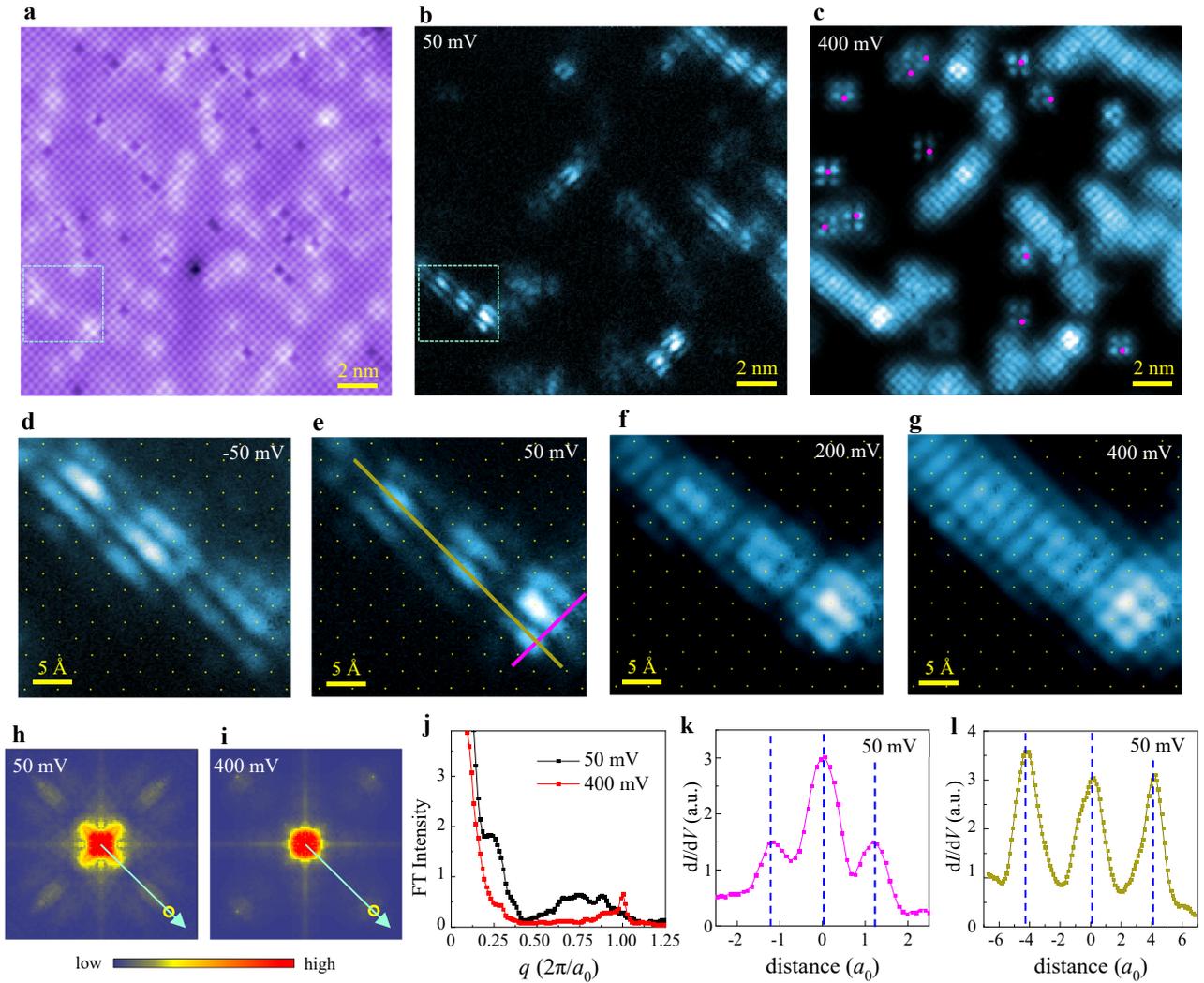

**Figure 2 | The stripe- and ladder-shaped molecular orbitals in the $p = 0.03$ sample taken at $T = 23$ K.** **a**, The topographic image of the $p = 0.03$ sample. **b-c,** The local DOS maps at $V_b = 50$ mV and 400 mV, respectively, in the same FOV as **a**. The low-energy electronic state in **b** consists of diluted stripe arrays immersed in the dark Mott insulator background, while the high-energy one in **c** has the ladder pattern and sporadic ZRSs with the dopants marked by magenta dots. **d-g,** Zoomed-in DOS maps on the area enclosed by the cyan box in **a** and **b**, which display the evolution from stripe- to ladder-shaped molecular orbitals with varied energies. **h-i,** The FT images of the DOS maps in **b** and **c**. The lattice Bragg points are indicated by yellow circles and the cyan arrows are along the Cu-O bond direction. **j,** The FT intensity along the cyan arrows in **h-i** reveal a broad range of wavevectors corresponding to short-range periodic patterns in real space. **k-l,** The line profiles of the DOS map at 50 mV in **e** along the magenta and golden lines. The inter-stripe distance perpendicular to the array is around $1.2a_0$, while that along the array displays the $4a_0$ periodicity.

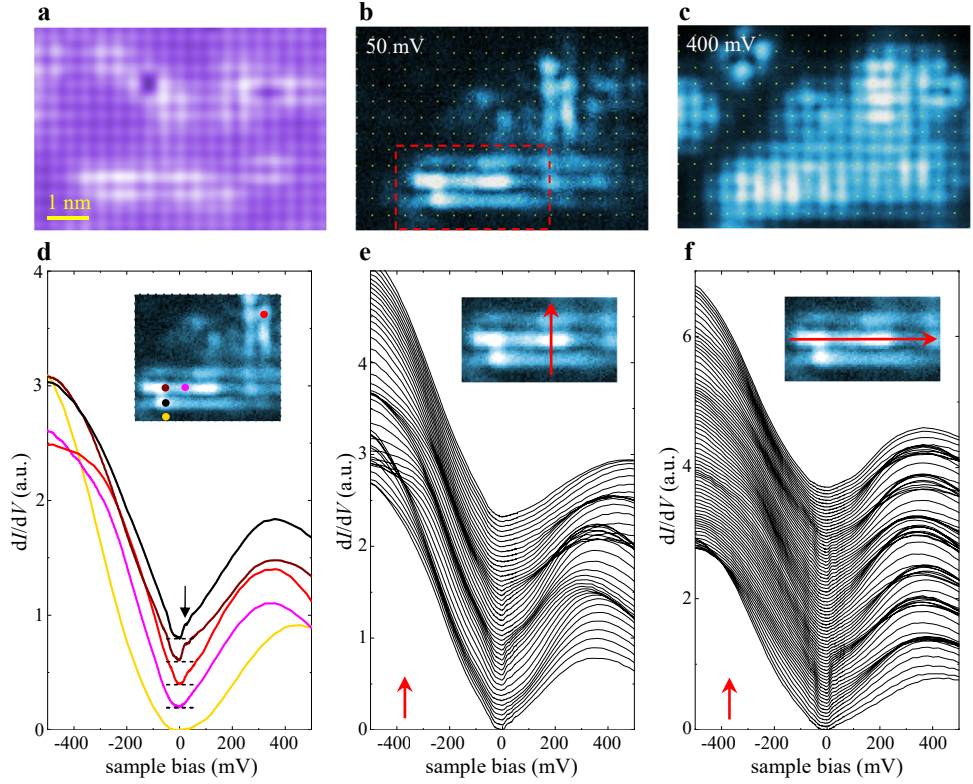

**Figure 3 | Low-energy electronic states in the *p* = 0.03 sample taken at *T* = 23 K. a**, Topographic image of an area with clusters of dopants. **b-c,** The DOS map of the same FOV taken at $V_b$ = 50 mV and 400 mV, displaying the stripe and ladder molecular orbitals, respectively. **d,** Typical low-energy spectra taken on the representative locations marked by the corresponding colored dots in the inset. There are preliminary gap-like features near $E_F$ on the bright stripes as indicated by the black arrow at energy scale ~15 meV. **e-f,** The d*I*/d*V* spectra taken along the arrows in the insets, corresponding to the area enclosed by the red dashed box in **b**. The two sets of data demonstrate that the gap-like features are closely tied to the stripe-like molecular orbitals, where the in-gap state peaks also occur at relatively lower energy.

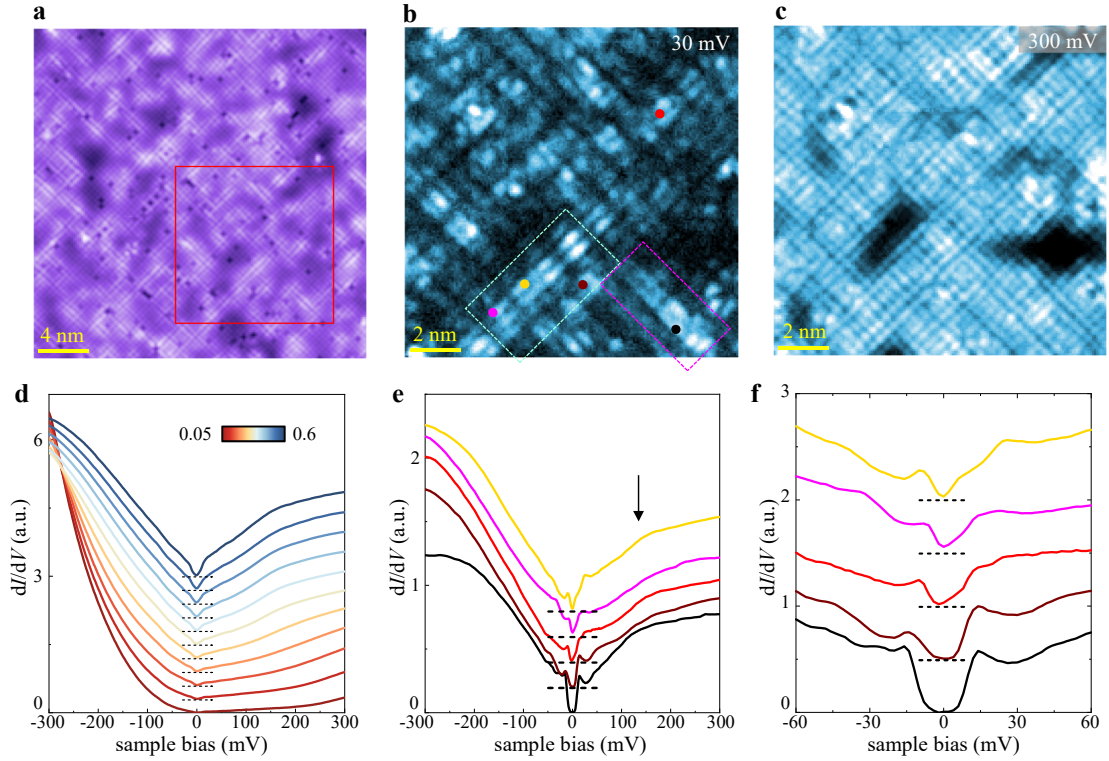

**Figure 4 | Topography, DOS maps, and electronic structure of the** $p = 0.07$ **sample taken at** $T = 5$ **K. a**, Topographic image of the $p = 0.7$ sample with pronounced stripy features along both Cu-O bond directions. **b-c,** The DOS maps at 30 mV and 300 mV in the red square of **a**, displaying denser occupancy of checkerboard with internal stripe and ladder patterns, respectively. **d,** The clustering analysis of the tunneling asymmetry at 300 mV, displaying the gradual emergence of low-energy gap near $E_F$ accompanied by the enhancement of pseudogap features. **e**, Typical d$I$/d$V$ spectra taken on the representative locations in **b** marked by corresponding colored dots. **f**, The low-energy d$I$/d$V$ spectra on the same spots as in **e**, showing the existence of both V-shaped and U-shaped gaps at different locations.

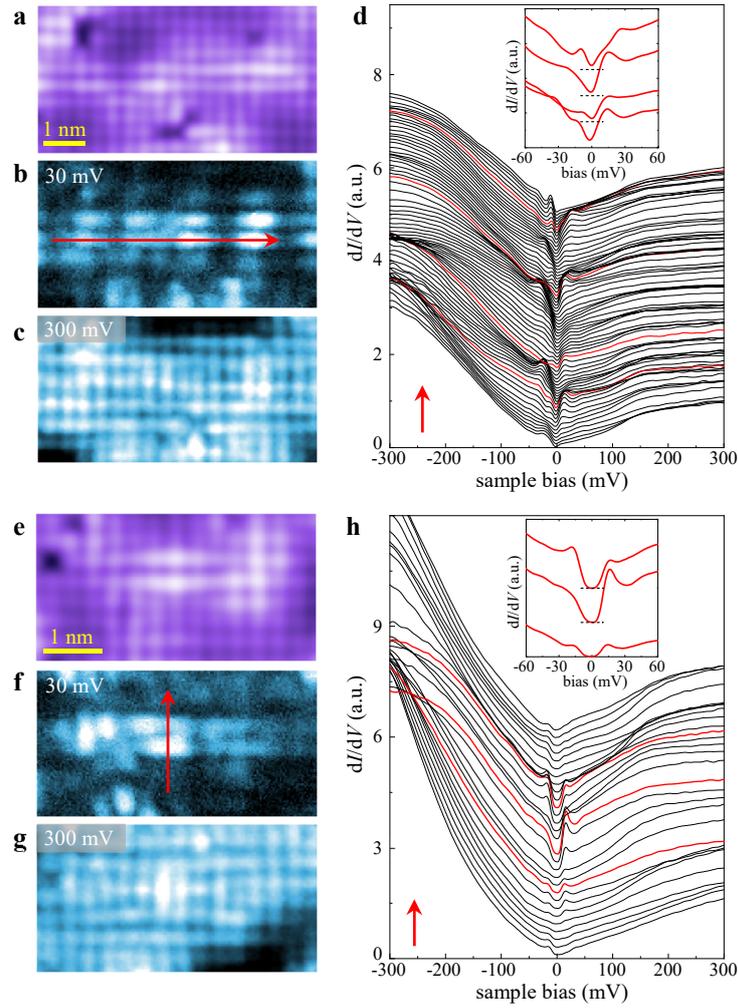

**Figure 5 | Distinct gap features in two different stripe arrays in the *p* = 0.07 sample. a,** Zoomed-in topography of a stripe array enclosed by the cyan dashed box in Fig. 4b. **b-c** The DOS maps taken at $V_b$ = 30 mV and 300 mV, respectively, revealing the stripe and ladder molecular orbitals that are highly similar to Fig. 3b-c for the *p* = 0.03 sample. **d,** The d*I*/d*V* spectra along the stripe direction as marked by the red arrow in **b**, displaying the nearly periodic modulation of electronic states. The inset shows the spectra on the bright stripes, which exhibit well-defined V-shaped gap reminiscent of the *d*-wave SC gap. **e-h,** The same set of measurements on the other stripe array enclosed by the magenta box in Fig. 4b. There is also a close correlation between the low-energy electronic states and the stripe-like molecular orbital, but here the small gap has a nodeless U-shape distinctly different from that expected for *d*-wave symmetry.

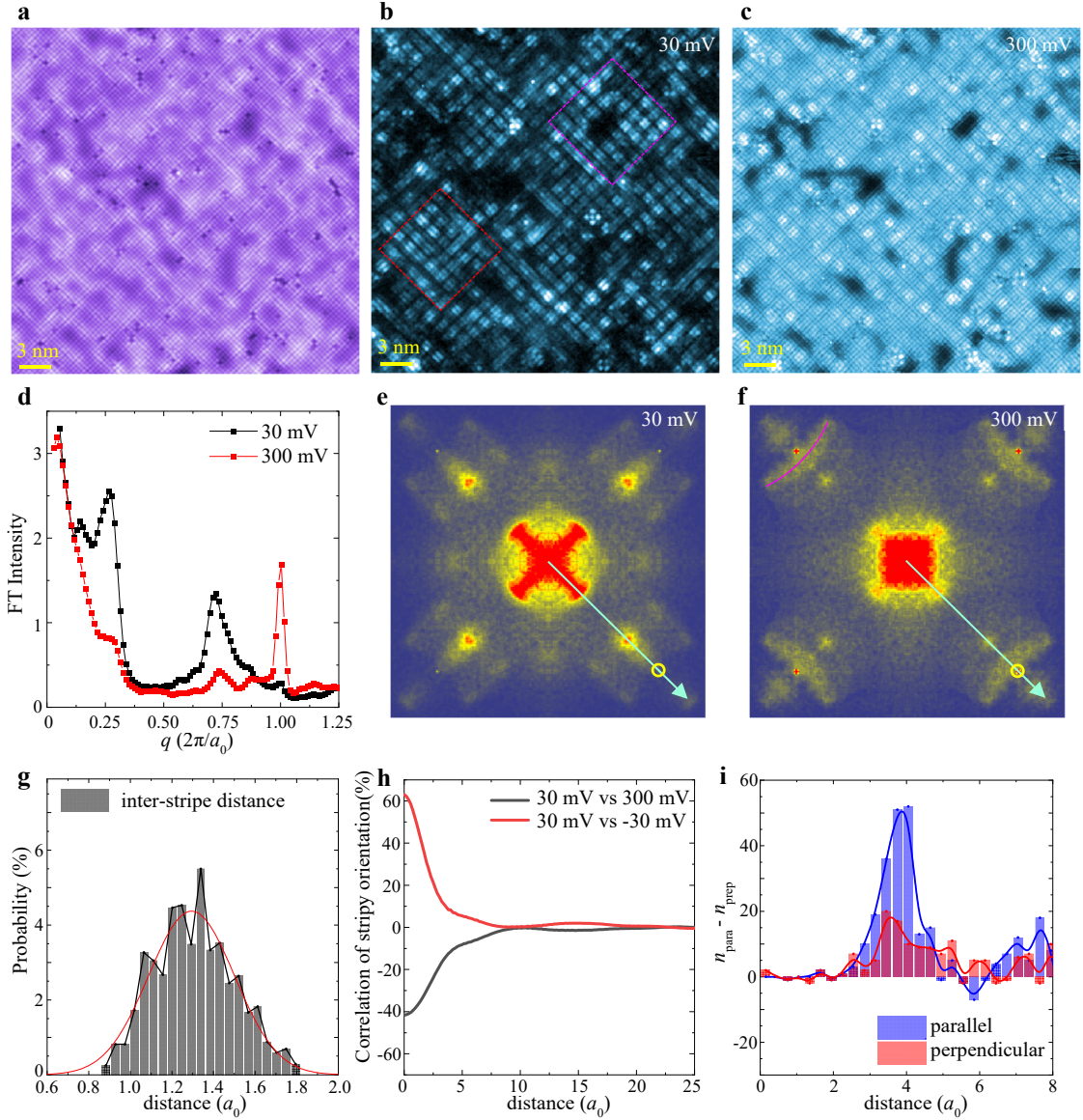

**Figure 6 | Real-space electronic structure of the SC $p$ = 0.10 sample taken at $T$ = 5 K. a,** Topographic image of the $p$ = 0.10 sample. **b-c,** DOS maps taken at $V_b$ = 30 mV and 300 mV in the same FOV as **a**, revealing the packed plaquettes of stripe and ladder molecular orbitals, respectively. **d,** The FT intensity along the cyan arrows in **e** and **f**. **e-f,** The FT images of **b** and **c**. The yellow circles indicate the lattice Bragg peaks. The map at 30 mV displays the renowned checkerboard order wavevector around $0.28 \times 2\pi/a_0$ and a broad range of incommensurate wavevectors around $0.72 \times 2\pi/a_0$, while the map at 300 mV is dominated by modulations with lattice periodicity. **g,** Distribution of the inter-stripe distance within each plaquette with a gaussian fit, which centers around $1.3a_0$. **h,** The positive correlation of local orientational intensity between stripes at positive and negative energies, and the negative correlation with the high-energy ladder. **i,** The histograms of inter-plaquette correlation for directions parallel and perpendicular to the stripes, respectively. They both peak at a distance slightly smaller than $4a_0$, but the parallel one is much more pronounced.

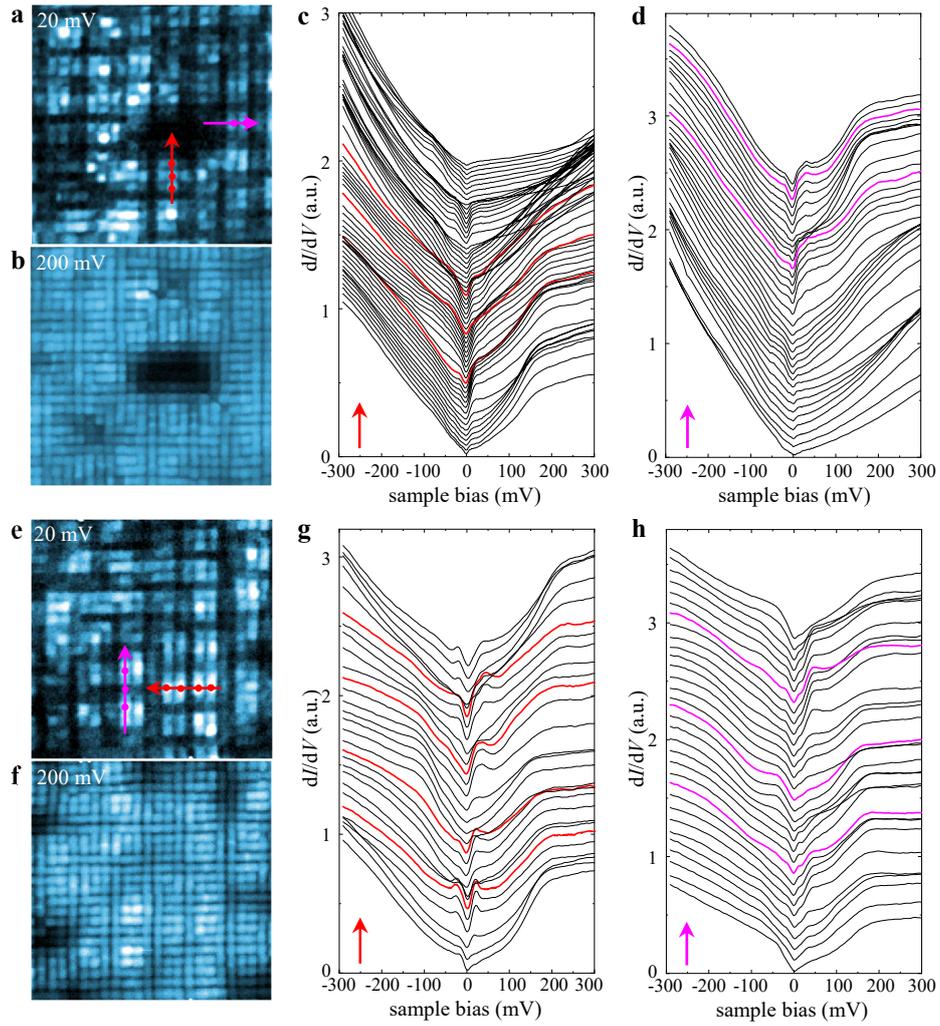

**Figure 7 | Spatial distribution of molecular orbitals and SC gap features on the $p = 0.10$ sample. a-b,** Zoomed-in DOS maps on the magenta dashed box in Fig. 6b taken at $V_b$ = 20 mV and 200 mV, respectively. The stripe and ladder molecular orbitals become densely packed plaquettes, with a regular dark pit with size around $4a_0 \times 8a_0$. There is no low energy DOS in the dark pit, and the topography in Fig. 6a shows the absence of dopants in this small region. **c-d,** The d$I$/d$V$ spectra taken along the arrows with corresponding colors in **a**. The V-shaped SC gap is more pronounced on the bright stripes and vanishes on the dark pit. **e-h,** The same sets of data as **a-d**, but on a region (red dashed box in Fig. 6b) with more regular checkerboard that appears like a mosaic of plaquettes with slightly different sizes and shapes. The d$I$/d$V$ spectra exhibit short-range periodic modulations with the most pronounced SC properties on the bright stripes.